\documentclass[10pt,journal,letterpaper,compsoc]{IEEEtran}

\usepackage{graphicx}
\begin{document}
%
\title{Remote Control of Mobile Devices in \\ Android Platform}

\author{Angel Gonzalez Villan, and Josep Jorba \\
Universitat Oberta de Catalunya \\
Barcelona, Spain \\
\{angonzalez, jjorbae\} @uoc.edu 

}


\IEEEcompsoctitleabstractindextext{%
\begin{abstract} 
Remote control systems are a very useful element to control and monitor devices quickly and easily. This paper proposes a new architecture for remote control of Android mobile devices, analyzing the different alternatives and seeking the optimal solution in each case. 
Although the area of remote control, in case of mobile devices, is little explored, it may provide important advantages for testing software and hardware developments in several real devices. It can also allow an efficient management of various devices of different types for performing different tasks, related for example to security or forensic tasks. 

The main idea behind the proposed architecture was the design of a system to use it as a platform which provides the services needed to perform remote control of mobile devices.
As a result of this research, a proof of concept was implemented. An Android application running a group of server programs on the device, connected to the network or USB interface, depending on availability. This servers can be controlled through a small client written in Java and runnable both on desktop and web systems.
\end{abstract}

\begin{keywords}
Android, Remote Control, VNC, Java, Mobile devices, Security solutions for mobile devices, Remote visualization
\end{keywords}
}

\maketitle

\IEEEdisplaynotcompsoctitleabstractindextext

%
\IEEEpeerreviewmaketitle

\section{Introduction}
\IEEEPARstart{T}{he} growing popularity and spread of smartphones has changed the design of computer systems as they were known in recent years. Technological developments have enabled the creation of mobile devices with technical features previously only conceived in PC architectures or similar devices. With this evolution comes the need to integrate these devices with others, as shown by Albrecht Schmidt and Dominik Bial[26], so they can take actions and monitor interaction on mobile devices, as is exposed . To this end, this paper proposes and analyzes different architectural approaches for the implementation of remote control systems of mobile devices using the Android software stack [1, 2].

Due to the spread of new 3G networks and the convergence between wireless network and wired network, performing tasks on mobile devices is easier and faster. Because of the easy to access of a network, previous protocols such as WAP were replaced by the direct connection to the network. This change involves that remote systems are not limited by network protocols with limited features [3].

In addition to enable the execution of tasks remotely, it is intended that the architecture allows performing software management tasks on the device and forensics tasks [4,5] in order to analyze the current state of the system and analyze the traces of previous executions to analyze the state of the device.
To perform communication between mobile devices and control equipment wired connections (using USB communication) and wireless connections are used. Communication standards will be analyzed with the aim of establishing a stable, optimal and safe communication.

An important consideration in designing the architecture of remote control of devices is the security within the system [6]. The device will exchange personal information of the user and the operations on the mobile device will be carried out remotely. It therefore must be ensured that no external element is able to both access the data exchanged, and to take control of external access.

Some related work is provided in Section 2. In section 3, a detailed analysis of remote visualization systems is presented. In section 4 is shown a connectivity assessment where is studied the different ways of connectivity. Section 5 presents the security considerations that should have a system of this type. Section 6 proposes a new remote control architecture. Section 7 shows the experiments performed and the results obtained. And finally it is concluded in section 8.

\section{Related work}
\IEEEPARstart{I}{n} the scope of remote control there are several projects and initiatives designed to allow remote control between devices. Although most of the architectures have the objective of control remotely PCs, there are some initiatives that aim to control mobile devices.

Remote control architectures offered by manufacturers cover only a part of the features required for an effective use, and usually are designed as internal solutions.

For instance, one manufacturer of Android devices, as Samsung, has a tool called Kies [8] (a commercial software), that allows the user to upgrade the firmware, control the contacts, music, photos and videos and control the file system. But it does not allow controlling applications, processes, services, etc.

Other aspect to be considered is the remote visualization mechanisms that is useful for achieve a remote display of the devices. The most popular system designed to perform remote control of devices is Virtual Networking Computing [9]. There are a large number of implementations to this solution including applied to Android software stack. It has an open protocol and it is widely deployed in the open source community. This solution adapts very well to provide part of the functionality of the architecture, and it will be studied further. 

Skurski and Swiercz [7] propose a control system based on VNC for Symbian OS smartphones. This system was designed to improve application testing systems in mobile devices due to the lack of resources in mobile devices and the high cost of test environments. Also the solution proposed could be used to perform remote configuration.

As part of the Android platform exists the Android Debug Bridge (ADB) protocol [10] to provide debug functionality on devices. The platform integrates this protocol and it offers a service of server when is configured on the device. To manage the communication with this protocol, the Android Development Kit [11] offers the ADB Client tool [10]. Some of the features of this tool are the installation and de-installation of applications, downloading and uploading files, opening a shell console, starting applications, etc.

The development of the Android platform is constantly evolving and therefore the features it offers are continuously expanding. For example, in the most extended versions of the platform like Android 2.3.3 the USB Host Support [12] feature is not available, but it's included in the new Android 4 (Ice Cream Sandwitch). The USB Host Support will allow attaching different devices such as keyboards, mice, etc. Also it is supposed that it will allow native control of the USB interface, avoiding the mandatory control of the Virtual Machine over the interface. This improvement will significantly increase the communication performance and offer a broader range of features.

This paper focuses on the control of Android platforms. This is an open platform that allows to use other technologies (also open). In addition, Android platform allow the development of new ideas easily and test them with a set of open standards [13]. The prototype generated as implementation of the proposed architecture will be provided also as free software. According to data released by Nielsen [14], half of the consumers who recently purchased a smartphone chose an Android smartphone.

Due to the lack of initiatives that provide complete remote control architecture oriented to open platforms, this paper presents a proposal that covers this area of interest. The proposed platform is open, flexible and scalable. Its use by monitoring teams of mobile devices implies an improvement in performance of management and control tasks.

\section{Remote Visualization}
\IEEEPARstart{T}{he} graphical visualization subject represents the most important aspect in a remote control system when a human supervisor is controlling the system. To achieve proper control of a mobile device supervisor need to watch the results of his actions and the best way is the graphical visualization. The display should be smooth and have a good quality to provide the best benefits to the user. The problems in the transfer of information condition this requirement, and therefore it requires manage resources efficiently.

\subsection{VNC System}

An alternative for displaying graphics and having graphic control of the device is the graphical desktop sharing system VNC (Virtual Network Computing). It offers a controlling functionality by using a graphical screen update from the remote device and capturing events like mouse or keystrokes [7]. VNC system is based on RFB (Remote Frame Buffer) [15] protocol to transmit all information between connected devices. 

The VNC system is compound by a server side and some thin clients that connect remotely to the server and send requests to the server to retrieve updates of the remote controlled device. This mechanism allows thin clients to keep light weight while the server offers them the data processed remotely. The server side tracks and encodes display updates, and the client side decodes and renders the updates received. The limiting factor of bandwidth is a problem due to the amount of that that is sent, above all because of the latency in the network [16]. There are VNC clients available for all of the most common systems, and it therefore can be considered a multiplatform system. This feature, along with the power of the protocol used by VNC, establishes it as a great choice for incorporation in the proposed architecture.

\subsubsection*{Encodings Study}

The way of working of the RFB protocol consists of responding to a request from the client about a specific on-screen rectangle and the server sends an update consisting of an encoding o the variation between the moment of the request and the last time the client requested data about this rectangle [16]. This action implies a high consumption of bandwidth in sending information, with the consequent delay in the process. In order to solve this problem different encodings have been developed. Encoding refers to how a rectangle of pixel data will be sent. Every rectangle of pixel data is prefixed by a header giving the position of the rectangle on the screen, the width and height of the rectangle, and an encoding type. This encoding type specifies the encoding of the pixel data. The data itself then follows using the specified encoding [16]. These encodings are methods to determine the most efficient way to transfer graphical information. Adding new encodings developed by third parties does not affect compatibility with VNC applications that do not contain that new encoding. When the client establishes communication with the server, both parts negotiate the encoding to use. If the client requires a non-existent encoding, the server will appropriate the next encoding available.

Below, the following VNC encoding will be studied: \textit{Raw}, \textit{RRE}, \textit{Hextile}, \textit{Zlib} and \textit{Tight}.
\begin{itemize}
\item[RAW] is the simplest encoding. It sends all graphical pixels to the client. This method must be supported by all clients. The process time used is minimal and the performance is very high when the server and the client are on the same machine. If the client is hosted in a remote device the performance is reduced due to the transfer of large amounts of data.
\item[RRE] (Rise-and-Run-length-Encoding) consists of grouping consecutive identical pixels in order to send only the information of one pixel and the number of replications. It is an effective method when large blocks of the same color exist, like in patterns. There is a variant of the protocol, called CORRE, which uses a maximum of 255x255 pixels to reduce the size of the packages.
\item[Hextile] divides the information into 16x16 items to be sent separately and in a predetermined order. The data of each item is encoded via Raw or RRE. The performance of this method increases as the network speed increases.
\item[Zlib] uses a mechanism to compress the information in order to reduce the size of the package as much as possible. The drawback is that it requires a great amount of CPU processing. This method is usually used when the VNC server does not work with the Tight encoding.
\item[Tight] uses the compression mechanism of Zlib to compress data, but it pre-processes the data in order to maximize the compression ratio and minimize CPU consumption. JPEG compression can be included for areas with many colors. This method is very efficient in all type of environments, even on slow networks.
\end{itemize}

\subsection{Native capture}
The VNC system is known for its high performance due to the power of its protocol, as was  previously mentioned, and for being compatible with almost all systems. But its use requires including a VNC server inside the device and also providing “superadmin” permissions to the Android application to use all the features. There is an alternative that allows the architecture to provide a remote visualization without including a VNC system, the native capture.

This system consists of access to the framebuffer device of the mobile device to extract bitmap data representing the screen surface image. Once the data has been extracted the system sends it to the clients that will display this data, showing the same information that visible to the mobile device user. Of course, this must implement different mechanisms to improve the performance, because when sending the raw information, the latency of the system will be high. However, it's clear that the remote visualization can be done implementing a native application and linking using JNI (Java Native Interface) with an Android application.

As will be discussed later, the ADB Client implements a method of retrieving the remote visualization using this system.

\subsection{Challenges}

The development of the new features in smartphones is changing the features of desktop-sharing systems. These improvements include new ways of interacting with devices and new sources of information. The integration of these elements is a challenge for the systems, which often are not prepared to support certain behaviors. Below are listed some of these challenges and solutions to try to solve them are offered.

\subsubsection*{Multi-touch}

For a long time, the methods to control a device were restricted to mice, keyboards and joysticks. But now, with multi-tactile screens this standard has changed. The Android system allows a new type of control, the multi-touch. This system allows performing various actions with the device at the same time, and with a link between them. For example, it is possible to create and imaginary line joining the two points where the user is pressing on the screen; or to give functionality to the action of bringing the points of pressure near to each other.

The standard method of remote interaction, as in a VNC system, does not support this type of behavior. Remote clients can emulate motion events and inject them into the device, but it is not possible send sequential events in order to be executed at the same time. This is a problem, because remote interaction does not advance with the new devices. To solve this problem, the system needs to be adapted to include all the possible scenarios.

On one hand, if the client has a device that allows the multi-touch system, the architecture should send the events to the remote system so that it could echo these multi-touch events. The actual system of sending clicks and keystrokes should be extended a new composite event. This new event could be a combination of various events sent as a package in order to be executed as a group in the remote system and not as single commands.

On the other hand, if the client does not have a multi-touch device, a local mechanism could be designed to allow him simulate a multi-touch event. The event designed in the preceding paragraph could also be used in this system. To create the event, the local visualization of the client must implement a system to emulate the multi-touch experience. For instance, a graphical system to establish points of pressure and their movements could be created to simulate the event. The user could design the event and customize every single aspect of the interaction.

\subsubsection*{Sensors}

Another way of interacting with the new mobile phones is through the use of sensors (movement, geolocation, closeness …). These sensors allow the user to get data from device's environment or to modify the behavior of the device. Using the remote control the physical interaction with the device is not possible, making it impossible to use those sensors. Also, the sensors that provide data can be interesting for the purposes of remote control architecture.

It would be necessary to establish some kind of mechanism as in the Argos emulator system[25], in which simulated data are submitted. To integrate the use of these sensors two systems need to be created. First, a mechanism to operate with the physical sensors should be created. This system could simulate, for example, the rotation of the device. Including this control, the dispatch of events will be completely covered. And second, the remote server must be able to send the information from the environmental sensors, allowing the client, for example, to request GPS data.

\section{Connectivity assessment}
\IEEEPARstart{T}{he} different methods to perform the connection determine the scope that the architecture can reach. This chapter presents a review of different types of possibilities establishing connectivity between the clients and the mobile device. The performance and security are conditioned by the communication channel used and therefore should be reviewed each environment to implement the necessary mechanisms.

\subsection{USB Interface}
The wired alternative to perform the communication between an Android device and a computer is the USB interface. Android smartphones are equipped by a MINI or MICRO USB port. This connection method usually can be used to share the filesystem, use the device as a remote media player or establish a connection to perform remote tasks.

At the time of writing of this article the feature of USB Host in Android platform is not distributed on Android Smarphones. The USB Host and Accessory feature is implemented on versions up to Android 3 Honeycomb and according to published news indicates that it could be backported to Android 2.3.4, but it has not yet been performed. The USB Host Support could allow the native communication without the mandatory control of the Android platform. This method will improve the performance and will increase the functionality available, allowing tasks directly on the integrated elements [12].

But until this feature is available, the only ways to establish USB communication are using the Android Debug Bridge capabilities or via local sockets. The ADB system allows USB connection to the device, having a server implemented in the kernel of the Android device by default. The capabilities of the server are limited by the developer's implementation and third parties cannot extended it. The communication over a local socket is very similar to a network communication. A client-server socket connection must be open, but the channel must be redirected to the USB interface. In this case the capabilities of the system are absolutely defined by the developer because he defines the handshaking protocol of the exchange [11].

The proposed solution in this environment is to mix both alternatives. On one hand the features served by the ADB system are implemented and tested, and can access to system resources. On the other hand the possibility of extending the existent features can be increased using local sockets. As we discuss in the next section, the wireless scenario requires a similar solution, so the need to implement a custom server is a core aspect of the architecture.

\subsection{Java sockets mechanisms}
Java offers some features like networking support, multithreading, multiplatform and portability that converts it in a suitable platform to be implemented in a distributed system. However, it has a problem of performance due to lack of efficient communication middleware, thus penalizing sync speed. Sockets are a low-level programming interface to manage networked communications in most of the network protocols. The Java Sockets can be used in every system that has Java Virtual Machine, but TCP/IP protocol limites it. In systems with high-speed networks this communication system is not appropriate because it does not make use of the benefits of the environment.

To create this communication system, the Android application will have a Server Socket running and awaiting client requests. To manage clients in a parallel manner, it will implement a system of threads that will serve each client concurrently, making use of the shared resources.

On the other hand, clients will have Java client sockets that will open communication with the server and will exchange messages with the server until the connection is closed [17].

Because the Java communication mechanism consists of the exchange of binary data; we will implement an interface of communication. This mechanism consists of defining a set of objects that contains a subset of objects representing both requests and responses. Thus both, client and server can use the channel of communication and understand the data they receive. To send objects over a network the system force us to save and restore these objects on both ends of the channel. To solve this problem, we will use Serializable Java Interface, making the process easier. Security issues regarding the exchange of objects will be studied in chapter V.

\subsection{Android Debug Bridge Client}
Android Debug Bridge Client is a tool existent in the Android SDK that allows to the developer to exchange data and requests with an Android device. These features can be adapted to be used inside a remote client. Originally, the ADB client was only an application to be used by a developer controlling a device or an emulator, but now its features have been packaged into a library, allowing its use in a Java application. This library has been created and given the name of “ddmlib” by the Android SDK Team.

It provides a great number of features for controling a device with USB. Some of the features implemented in this library are:

\begin{itemize}
\item Executing shell commands
\item Forwarding ports
\item Dumping device data
\item Managing the file system
\item Managing applications
\item Checking the state of the device
\end{itemize}

\subsection{Conclusion}
The choice between using wireless connection or a USB interface is complex because both have advantages and disadvantages. As each of the methods offers a number of complementary features, it was decided to use a mixed solution.

As the ADB tool has implemented a set of features, to avoid rewriting this part of the system, they will be available when the device is connected via USB. Functions not covered will be implemented in the application server to be accessible through the two types of connection. The mechanism consists of exchanging items that identify the requested operation from the client to the server. This object contains a code of operation and an associated message. The server reads the information of the object, performs the operation requested and sends back the results obtained. The client receives this data and shows it to the user.

Connectivity via USB is done via cable, thus offering better performance but requiring the availability of the device near the control equipment. Connection via network, offering less functionality however, allows more freedom in localization of devices. The system will provide flexibility in connection options to let the user choose the most appropriate method at all times.

As a consequence of using a VNC server, the connection with this part of the system is available via web browser. Most of the browsers can integrate a VNC client, so if a URL is entered, the browser connects to a VNC server, it launches the VNC client and establishes a connection. The VNC client integrated also allows graphical interaction if the server implements this feature. As a result of this feature, access to a limited set of features is available if there with only browser capability.

In this paper the connection via network and USB is contemplated, but the approach could be adapted to study the use of technologies such as bluetooth. Also, the method of remote access is ready to support different connection methods that may appear in the future and allow better communication. 

\section{Security Management}
\IEEEPARstart{T}{he} increasing number of attacks on mobile devices (especially on smartphones) along with the expanding use of the same make clear that these devices must implement security mechanisms [18].  Android Platform was designed to enable applications to run in specialized sandboxes, preventing the spread of the malware. The security between the applications and the system is applied at the process level by standard Linux facilities. Two applications cannot run in the same process unless the sharedUserID feature is used [19, 20].  Thus prevents a malicious application from accessing data from another application process. As part of the system uses native code, this code should be implemented following preventive measures because its implementation is beyond the control of the virtual machine [21].

The communication transports provides opportunities for malware to infiltrate a device. The system under analysis communicates through both cable and network and must ensure these connectivity methods. The USB protocol is entirely managed by software and the sub-protocol used in this system is the ADB. To use the system the device must have the USB debugging enabled. This architecture implements a mechanism of objects to define the commands. These objects are processed to avoid a man-in-the-middle attack that could alter the data exchanged.

In terms of connectivity through a network, the permission “INTERNET” must be granted to the application [11]. In the Android platform the applications need special permissions to open network connections by being members of relevant groups. These applications are allowed to create sockets. The system must have these permissions, so there must be assurance regarding preventing the execution of malicious code.

To ensure the system, we could use an encryption method based on passwords or certificates. With this system the clients will need valid authentication data to establish the connection with the server. Each device will have its own login method avoiding the connection of clients not authorized. Also, the data exchanged should be transferred encrypted using a secure method, avoiding the capture and decode of the objects exchanged. This method could use a system based in private and public keys ensuring the valid origin of the data and discarding the data sent by third elements with malicious purposes.

Therefore, any implementation of the architecture must implement security mechanisms in all aspects that could represent a risk to the system. Depending on the environment in which the architecture is implanted, the degree of security must adapt to ensure system security.

\section{Remote Control Architecture Proposal}
The architecture proposed in this paper consists of a remote control architecture of mobile devices on the Android platform based on a client / server model oriented to services. The server layer is performing the services of mobile device management and accepts the connection from different clients. The client layer, available from a remote device, performs the interaction between the control equipment and the monitored device. As can be observed in Figure 1, the architecture offers several types of connection to different clients in order to allow the remote control to all the users. The monitored devices can be aggruped in a server in order to offer the control of several devices available in the same infrastructure.

\begin{figure*}[htb]
	\centering
	\leavevmode
	\includegraphics[scale=0.35]{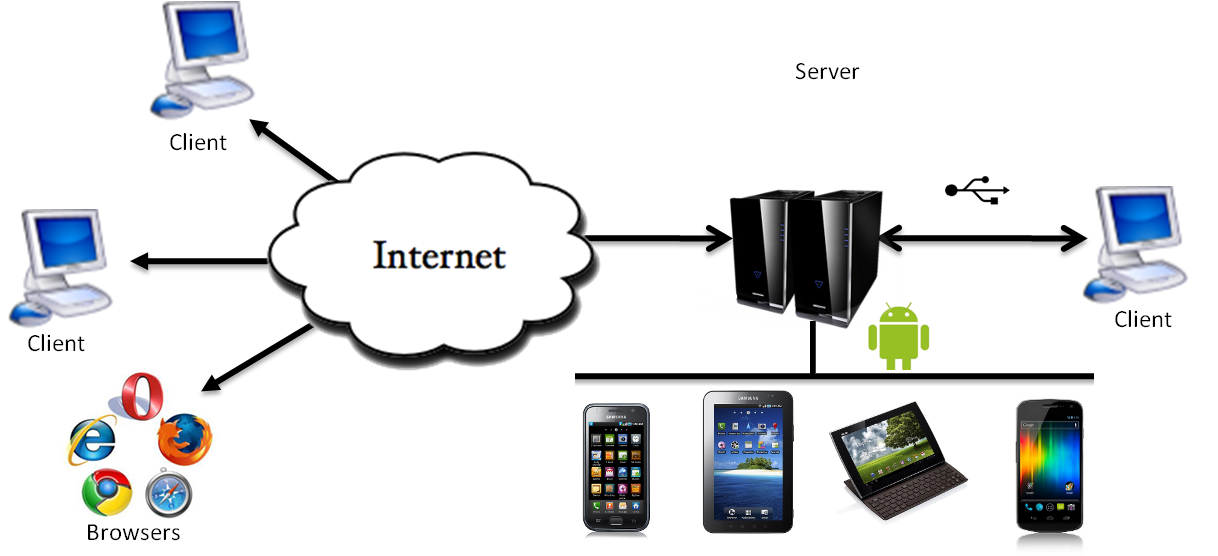}
	\caption{Remote Control Architecture Proposal}
\label{fig:architecture}
\end{figure*}

Below, the features that will implement the architecture will be listed and will be determined the chosen solution will be determined.

\subsection*{Remote visualization service}
The server layer offers a VNC server service to share graphic information to the client layer. We will use the Tight encoding in the VNC service. Using this configuration, the display will be smooth even when the network is slow. The client service requests for the connection parameters and establishes a connection. While the interaction is active, the client layer requests updates to the server to show the device's display. If the server layer does not support the VNC system, a raw display will be offered as an alternative.

\subsection*{Application management service}
Another service offered by the server layer is the centralized management of applications. The client layer can retrieve the information concerning to the applications and also modify the remote and modify the remote situation. Thus it would be possible to perform software updates on all monitored devices. 
For example, the client service could request the installation of an application and the server layer is responsible of installing the file sent by the client.

As an improvement this feature could be integrated with the Android Web Market, enabling search and downloading of applications.

\subsection*{Service and process management service}
At any moment, the device has a set of devices and processes running. Any of these tasks could be making a bad use of the resources or simply require its completion. The server service offers the information of these processes and services running. The client layer can manage these processes and services to achieve a particular behavior. All this management is accomplished by the server service, by virtue of having access to the processes and their life cycles.

\subsection*{Shell console service}
As a Linux system, when a more advanced control system is required, it is very useful to have a shell console. With this access point the control user could perform all the operations needed directly in the device. This feature involves a great responsibility because any failure or erasure can be fatal for the device.

\subsection*{File system management service}
A common need in working with mobile devices is the exchange of files between both systems. The user could need a file generated by the mobile to process or use it in another system. Using this service it is possible to do it without any action from the device. Furthermore, the client could inject files into the device, for example, to upgrade the firmware, make some data available to users, etc. The server will offer the information of the remote filesystem and will allow requesting operations of getting, putting or removing files.

\subsection*{Device status service}
In order to quickly check the status of a device, the server provides the general information of the device to the clients. Thus the control user could determine if any of the devices require immediate attention. For example, the server could send an alert to the clients to notify the clients a problem or to request interaction.

\subsection*{Sensor data service}
Devices have a variety of sensors that provide environmental information. This information is very useful to determinate the device's status, its location and the presence of a user with the device. The server will offer this information when it is available. These data are only readable because the facility to change the status of the sensors in a remote way is not considered.

\subsection*{Firmware management service}
A useful feature when performing the control of several devices is the firmware management thereof. This functionality is critical in the system because, if not done properly, it is possible to lock the device. However,  some manufacturers offer alternative methods to modify the firmware. As the goal of the architecture is to embrace the Android devices offering the common method will proposed. This method consists of putting the file of the firmware upgrade inside the SD card and performing the upgrade. To performs this upgrade it is necessary to boot the device in the “recover mode” and apply the installation. As this process requires user interaction in that the device must be restarted with the consequent loss of communication, the process cannot be done completely remotely. As a temporary solution until there is a common method that allows remote mode, we propose a method to copy the firmware on the device and provide the user with instructions to complete the process.

\section{Experiment and analysis}
\subsection*{Performance study of VNC system}

\IEEEPARstart{T}{e} section 3 the different encodings available were analyzed, and now a typical test scenario of workload for remote visualization has been designed. The goal of this task is to determine which of the encodings under analysis is the most suitable for use within the proposed architecture. Thus, this analysis can guide developers in improving the performance of their remote visualization system.

We have developed a scenario of activities to be performed in the same way, with the 5 different protocols. The list of activities is as follows:

\begin{itemize}
\item At the start, the mobile device is in the HOME.
\item Open the browser.
\item Wait 3 seconds.
\item Open the music player.
\item Wait 3 seconds.
\item Return to the HOME
\item End the benchmark.
\end{itemize}

This scenario generates a display of about 10 seconds in which the device's screen is constantly changing. Thus the system is submitted to a limit situation in terms of visualization of the device. An automatic test case has been used to achieve the some execution in every case, since a manual execution would not provide completely reliable data.

This benchmark was executed in two different environments: device connected via USB and device connected via WIFI in a local network.

Device connected via USB:

\begin{table}
\small\addtolength{\tabcolsep}{-3pt}
\caption{Results of the analysis via USB}
  \begin{tabular}{ | c | c | c | c | c | c | }
    \hline
     & RAW & RRE & Hextile & Zlib & Tight \\ \hline
    Updates & 30 & 34 & 55 & 31 & 62 \\ \hline
    Updates/second & 2,23 & 2,17 & 3,57 & 1,62 & 3,6 \\ \hline
    Rectangles received & 30 & 34 & 55 & 372 & 1300 \\ \hline
    Data captured (MB) & 42,48 & 49,80 & 79,10 & 45,41 & 90,82 \\ \hline
    Data compressed (MB) & 42,48 & 32,49 & 37,28 & 7,40 & 3,94 \\ \hline
    Compression ratio & 1 & 1,53 & 2,12 & 6,13 & 23,06 \\
    \hline
  \end{tabular}
\end{table}
\begin{figure}[htb]
	\begin{center}
	\leavevmode
	\includegraphics[scale=0.7]{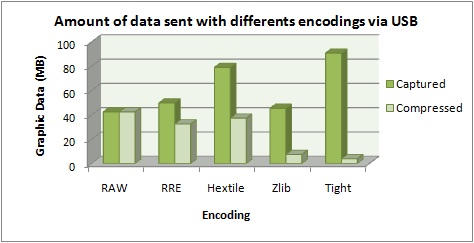}
	\end{center}
	\caption{Amount of data sent with different encodings via USB}
\label{fig:data_encodings_usb}
\end{figure}

As can be observed in the Table 1, Hextile and Tight protocols offer a great number of updates to the client. But Tight shows its power, being the protocol that sends the greatest amount of information over the wire with the lowest size of data exchanged. Zlib protocol also exchanges a small amount of data due to its compression mechanism, but the quality of the graphical information is significantly lower. Also, as Fig. 2 shows, Hextile has the problem of exchanging almost ten times the data exchanged by Tight, and thereby occupying the bandwidth.

Device connected via WIFI in a local network:

\begin{table}
\caption{Results of the analysis via WIFI}
\small\addtolength{\tabcolsep}{-3pt}
  \begin{tabular}{ | c | c | c | c | c | }
    \hline
     & RAW & Hextile & Zlib & Tight \\ \hline
    Updates & 8 & 20 & 68 & 71 \\ \hline
    Updates/second & 0,36 & 0,85 & 1,77 & 3,54 \\ \hline
    Rectangles received & 8 & 22 & 808 & 868 \\ \hline
    Data captured (MB) & 10,25 & 27,83 & 98,27 & 104,00 \\ \hline
    Data compressed (MB) & 10,25 & 5,73 & 8,98 & 3,62 \\ \hline
    Compression ratio & 1 & 4,86 & 10,95 & 28,77 \\
    \hline
  \end{tabular}
\end{table}

\begin{figure}[htb]
	\centering
	\leavevmode
	\includegraphics[scale=0.7]{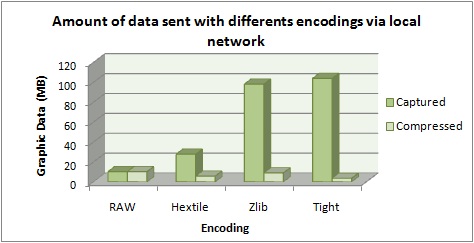}
	\caption{Amount of data sent with different encodings via WIFI}
\label{fig:data_encodings_wifi}
\end{figure}

Using this type of wireless connection, the VNC system sends less data due to the slower connection speed. The number of updates is lower and consequently, less is captured. In this scenario, the compression implemented by the encoding allows the system to send more updates regardless of the speed problem, as shown in Fig. 3. The compression ratio of Zlib and Tight encoding (10,95 and 28,77 respectively) converts this encoding in the more suitable option to be used in this type of scenario, as can be observed in Table 2.

As a result of the study, it is clear that the Tight encoding offers the best performance in any of the scenarios. It sends an average of 3,5 updates per second, allowing a smooth display that will offer a correct remote visualization to the user. Also it consumes a small amount of data through the communication channel.

\subsection*{Prototype implementation of the architecture: Android Remote}

As an approach to the use of the architecture proposal has been deployed a prototype project called “Android Remote”. This prototype of the architectural proposal consists of an application server and a thin client that shows the power of the proposal. The prototype project is licensed under the LGPL license to free distribution and modification of the project. The project can be used as a base for a more complete implementation.

The server application has been implemented as an Android application, designed to allow the user to activate the services provided by the server layer. The application will be available on Android Market to be downloaded and installed. It requires root permissions to start the remote visualization service based on VNC. Also the user will be notified when a client connects to the system, controlling the remote clients using the server application.
To design and implement the server application has been based on the free software application “droid VNC server” [22]. Fig. 4 shows the interface of the Android Remote server application.

\begin{figure}[htb]
	\centering
	\leavevmode
	\includegraphics[scale=0.3]{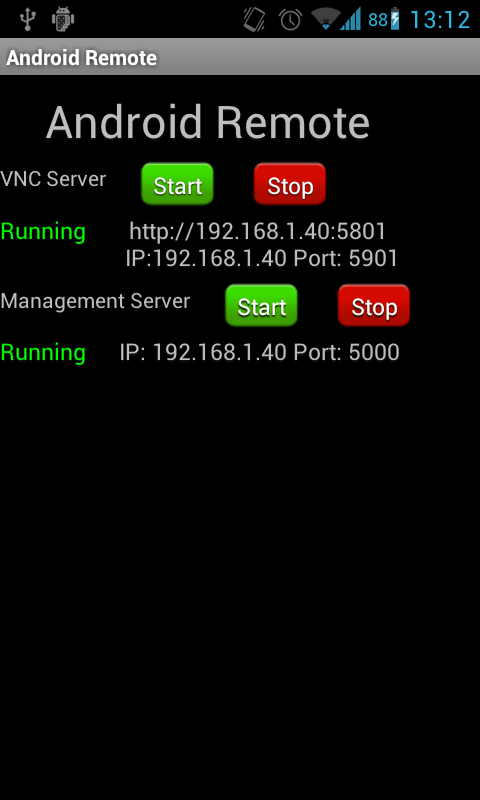}
	\caption{Android Remote Server}
\label{fig:server}
\end{figure}

To apply the features of the client layer, a thin client based on Java has been implemented. This client is able to connect to a remote device using a USB or network connection setting connection parameters. The client is control panel where is available a method to request some of the services offered by the server layer. The main purpose of the client is providing proof of concept to use the services. To design and implement the client application has been based on the free software application “AndroidScreencast” [23].

Some of the features provided are the following:

Remote display of the device using the VNC system. This feature allows watch the device and control it graphically. The control user can use the device directly. As shown in Fig. 5, the client offers some features to interact with the device.

\begin{figure}[htb]
	\centering
	\leavevmode
	\includegraphics[scale=0.6]{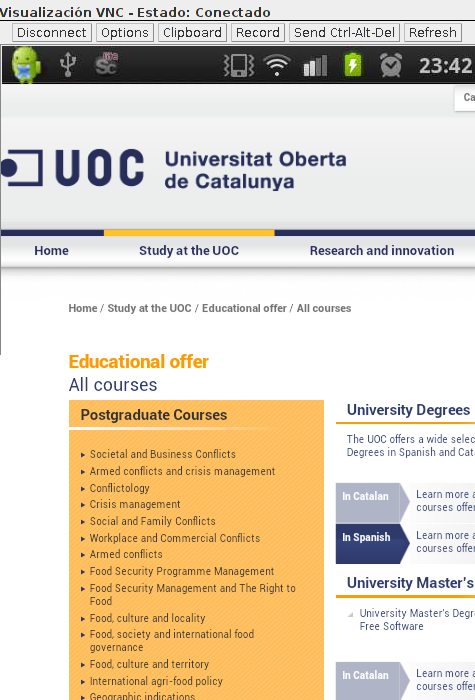}
	\caption{Remote display of the device}
\label{fig:remote_display}
\end{figure}

Application management of the software installed in the device. Control user lists the applications installed and running to check the applications used in the device. Fig. 6 and 7 show the lists of applications running and installed, respectively.

\begin{figure}[htb]
	\centering
	\leavevmode
	\includegraphics[scale=0.6]{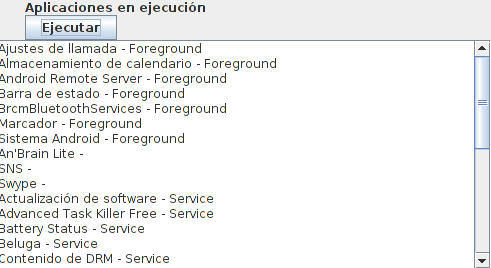}
	\caption{Running Applications}
\label{fig:applications_list}
\end{figure}
\begin{figure}[htb]
	\centering
	\leavevmode
	\includegraphics[scale=0.5]{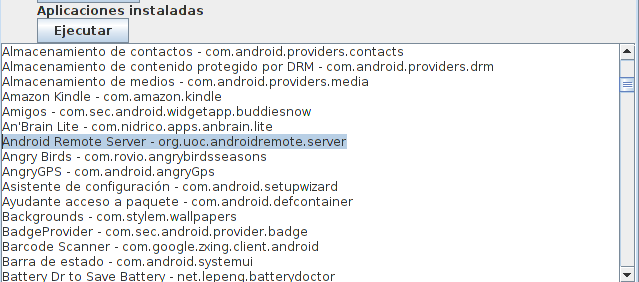}
	\caption{Installed Applications}
\label{fig:applications_installed_list}
\end{figure}

Filesystem management to access to the folders and files stored in the device. The server provides access to the information of the filesystem. The client displays a graphical and hierarchical view of the folders to allow the control user to check the data stored. Fig. 8 shows the explored opened by the client application to explore the remote filesystem.

\begin{figure}[htb]
	\centering
	\leavevmode
	\includegraphics[scale=0.5]{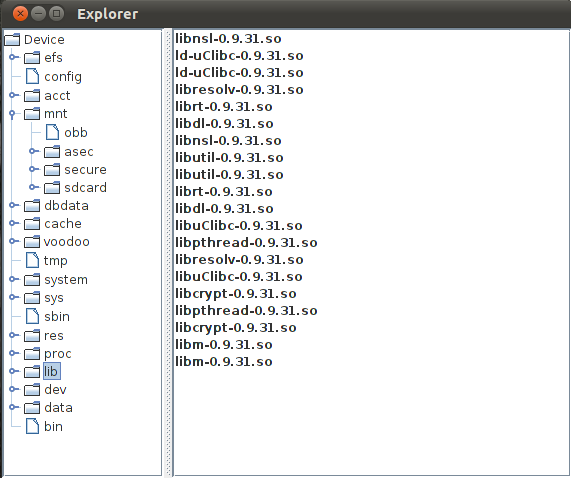}
	\caption{Remote filesysem}
\label{fig:filesystem}
\end{figure}

\section{Conclusion}
\IEEEPARstart{W}{e} have introduced an architecture to perform the remote control of Android devices. In particular, we have analyzed different alternatives to perform the most relevant aspects, determining their strengths and weakness. As a result of the research process, the main features have been identified, and the architecture should offer the most optimal procedures to carry out the exchange of data.

The remote visualization will be delegated to a VNC system only if the device allow the use of this software. Otherwise, the system will have to implement a native method to capture the graphics.

To perform the connection two methods are available, USB interface or Socket networking. The connection via USB will take advantage of the networking features in addition to the features implemented only to the USB connection. Thus, the USB connection will be complete. However, the Socket networking connection does not contain the exclusive features of the USB connection, but this problem can be solved developing the server features offered by the Android Debug Bridge.

Lastly, the system must implement a security system based in some security mechanism like encryption, password, certificates, etc. in order to ensure the data exchange and to avoid intrusions.

Thus, in future works, we will continue the implementation of the proposed architecture and will make the research to integrate the challenges defined. The multi-touch events and the sensor data injection represent an important part of the remote control and should be included.
As an expansion of the architecture, it has been implemented the concept of a mobile laboratory for development with multiple smartphones or tablets. This mechanism can offer security, observability, dangerousness, accessibility and availability to the remote laboratory paradigm[27].

The results obtained have been applied in a software prototype that shows the capabilities of the architecture. This prototype is available under a free software license as a complement of the paper. The server is an Android Project available in Google Code (http://code.google.com/p/androidremote-server/). The thin client implemented as Java project also available in Google Code (http://code.google.com/p/androidremote-client/).






\end{document}